\documentclass{IEEEtran4PSCC}

\ifCLASSINFOpdf
   \usepackage[pdftex]{graphicx}
\else
   \usepackage[dvips]{graphicx}
\fi
\usepackage[cmex10]{amsmath}

\usepackage{algorithm}
\usepackage{algorithmicx}
\usepackage{algpseudocode}
\usepackage{cite}
\usepackage{comment}
\usepackage{bbm}
\usepackage{hyperref}
\usepackage{eqparbox}

\usepackage{color}
\usepackage{mathtools}
\usepackage{amsfonts}
\usepackage{import}
\usepackage{tikz}
\usepackage{subfigure}
\usepackage{enumitem}
\hypersetup{
    colorlinks=true,
    linkcolor=black,
    filecolor=magenta,      
    urlcolor=black,
    pdftitle={},
    pdfpagemode=FullScreen,
    citecolor = black,
    }
\usepackage{amsthm,amsmath,amssymb}
\allowdisplaybreaks

\newtheorem{theorem}{Theorem}

\newtheorem{remark}{Remark}

\DeclareMathOperator*{\argmax}{arg\,max}
\usepackage{multirow}
\usepackage{graphicx, pifont} 
\usetikzlibrary{arrows.meta}
\usetikzlibrary{positioning,fit,backgrounds}
\usetikzlibrary{decorations.markings}

\usepackage{tikz}
\usepackage{tikz-network}
\usepackage{ifdraft}
\algnewcommand\Output{\item[\textbf{Output:}]}

\newtheoremstyle{mystyle}
  {\topsep}   
  {\topsep}   
  {\itshape}          
  {0pt}       
  {\bfseries} 
  {:}         
  {5pt plus 1pt minus 1pt} 
  {}          
\newtheorem{corollary}{Corollary}[theorem]
\theoremstyle{mystyle}

\newtheorem*{pov}{Probability of Violation}
\usepackage{cite}
\usepackage{amsmath,amssymb,amsfonts}
\usepackage{graphicx}
\usepackage{textcomp}
\usepackage{xcolor}
\usepackage{float}
\usepackage{mathtools}
\usepackage{booktabs}
\usepackage{multirow}
\usepackage{comment}    
\usepackage{tikz}
\usepackage{tikz-network}
\usetikzlibrary{arrows.meta}
\usetikzlibrary{positioning,fit,backgrounds}
\usetikzlibrary{decorations.markings}

\usepackage{multicol}

\makeatletter
\let\old@ps@headings\ps@headings
\let\old@ps@IEEEtitlepagestyle\ps@IEEEtitlepagestyle
\def\psccfooter#1{%
    \def\ps@headings{%
        \old@ps@headings%
        \def\@oddfoot{\strut\hfill#1\hfill\strut}%
        \def\@evenfoot{\strut\hfill#1\hfill\strut}%
    }%
    \def\ps@IEEEtitlepagestyle{%
        \old@ps@IEEEtitlepagestyle%
        \def\@oddfoot{\strut\hfill#1\hfill\strut}%
        \def\@evenfoot{\strut\hfill#1\hfill\strut}%
    }%
    \ps@headings%
}
\makeatother

\psccfooter{%
        \parbox{\textwidth}{\hrulefill \\ \small{24th Power Systems Computation Conference} \hfill \begin{minipage}{0.2\textwidth}\centering \vspace*{4pt} \includegraphics[scale=0.06]{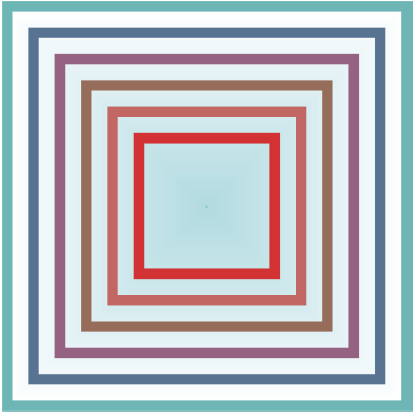}\\\small{PSCC 2026} \end{minipage} \hfill \small{Limassol, Cyprus --- June 8 -- June 12, 2026}}%
}

\begin{document}

\title{Learning Power Flow with Confidence: A Probabilistic Guarantee Framework for Voltage Risk}

\author{%
\IEEEauthorblockN{Parikshit Pareek\IEEEauthorrefmark{1},
Sidhant Misra\IEEEauthorrefmark{2},
Deepjyoti Deka\IEEEauthorrefmark{3}}
\IEEEauthorblockA{\IEEEauthorrefmark{1} Department of Electrical Engineering, Indian Institute of Technology Roorkee, India, pareek@ee.iitr.ac.in}
\IEEEauthorblockA{\IEEEauthorrefmark{2} Los Alamos National Laboratory, New Mexico, USA, sidhant@lanl.gov}
\IEEEauthorblockA{\IEEEauthorrefmark{3} MIT Energy Initiative, Boston, USA, deepj87@mit.edu
}
}

\vspace{-2em}

\maketitle

\begin{abstract}
The absence of formal performance guarantees in machine learning (ML) has limited its adoption for safety-critical power system applications, where confidence and interpretability are as vital as accuracy. In this work, we present a \emph{probabilistic guarantee} for power flow learning and voltage risk estimation, derived through the framework of Gaussian Process (GP) regression. Specifically, we establish a bound on the expected estimation error that connects the GP’s predictive variance to confidence in voltage risk estimates, ensuring statistical equivalence with Monte Carlo–based ACPF risk quantification. To enhance model learnability in the low-data regime, we first design the \emph{Vertex-Degree Kernel} (VDK), a topology-aware additive kernel that decomposes voltage–load interactions into local neighborhoods for efficient large-scale learning. Building on this, we introduce a \emph{network-swipe active learning} (AL) algorithm that adaptively samples informative operating points and provides a principled stopping criterion without requiring out-of-sample validation. Together, these developments mitigate the principal bottleneck of ML-based power flow—its lack of guaranteed reliability—by combining data efficiency with analytical assurance. Empirical evaluations across IEEE 118-, 500-, and 1354-bus systems confirm that the proposed VDK-GP achieves mean absolute voltage errors below 1E-03 p.u., reproduces Monte Carlo–level voltage risk estimates with 15$\times$ fewer ACPF computations, and achieves over 120$\times$ reduction in evaluation time while conservatively bounding violation probabilities.
\end{abstract}



\thanksto{This work was initiated when PP was a postdoctoral scholar and DD was a scientist with the Theoretical Division (T-5), Los Alamos National Laboratory, NM, USA. The authors acknowledge the funding provided by LANL’s Directed Research and Development (LDRD) projects: “High-Performance Artificial Intelligence" (20230771DI), the Artimis project, and The Department of Energy (DOE), USA under the Advanced grid Modeling (AGM) program. PP also acknowledges support from the ANRF PM Early Career Research Grant (ANRF/ECRG/2024/001962/ENS), received at IIT Roorkee. The research work conducted at Los Alamos National Laboratory is done under the auspices of the National Nuclear Security Administration of the U.S. Department of Energy under Contract No. 89233218CNA000001.}

\section{Introduction}
The increase in uncertain power sources and variable loads has made ensuring secure power system operation more challenging than before \cite{barry2022risk}. An important problem in this context is voltage risk assessment (VRA) that aims to quantify the likelihood of a bus voltage exceeding its operational limit due to uncertainty \cite{mccalley1999voltage}. The problem of VRA can also be viewed as uncertainty quantification (UQ) \cite{uq1} for the distribution of output voltage under uncertain load for a given operating condition. Computationally, performing VRA is a challenge as Alternating Current Power Flow (ACPF) equations are nonlinear and are not expressed as analytical (or closed-form) expressions of nodal voltages with bus load vector as input \cite{grainger2003power,pareek2021framework}. Instead, iterative methods such as the Newton-Raphson load flow (NRLF) must be employed which can lead to significant computational overhead since accurate VRA requires a large number of power flow samples.  On the other hand, the direct current approximation for PF \cite{stott2009dc,molzahnsurvey} neglects the voltage information, and therefore cannot be utilized to estimate voltage risk.

Recently, machine learning (ML) methods, in particular Deep Neural Networks (DNNs), have made significant progress as universal function approximators, especially in conjunction with the idea of physics-informed ML \cite{karniadakis2021physics,9743327}. They have also been explored for PF learning and UQ  \cite{gao2023physics,hu2020physics,chen2023physics}. The idea behind learning ACPF is that a fast PF solver can be used as an \textit{Oracle} to provide large number of voltage solutions for risk assessment \cite{gao2023physics}. However, DNNs require extremely large number of samples to learn the ACPF approximator. For instance, over $10,000$ training samples are used to learn voltage solutions for the 118-Bus system in \cite{gao2023physics,hu2020physics,chen2023physics}.



\begin{figure}[t]
    \centering
    \includegraphics[width=0.8\columnwidth]{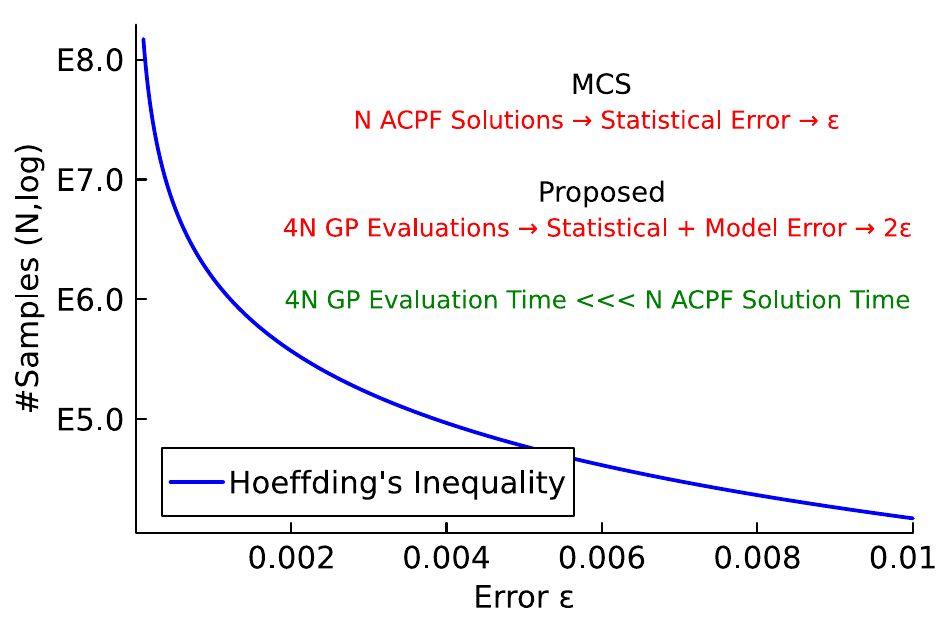}
    \vspace{-1em}
    \caption{Relationship between required number of samples $N$ and estimation error $\varepsilon$ for MCS-based voltage risk assessment (VRA). The proposed approach replaces these $N$ ACPF solutions with $4N$ GP model evaluations to achieve an error of comparable order. The key advantage arises from the GP’s closed-form evaluation, where the time per GP evaluation $T_{\text{GP}}$ is significantly smaller than the time per ACPF solution $T_{\text{solve}}$, leading to an overall speedup on the order of $T_{\text{solve}} / T_{\text{GP}} \gg 1$. Note that learning the GP surrogate requires only $N_{\text{train}} \ll N$, thus reducing the overall computational burden.}
    \vspace{-1.5em}
    \label{fig:hoeff}
\end{figure}

This work is built to address the need of having probabilistic grantee in voltage risk estimation, without exponential time complexity with respect to system size. For this purpose we use Gaussian process (GP) for modeling voltage-injection characteristic and use it for VRA. GP learning \cite{williams2006gaussian} is a versatile probabilistic method for function estimation that enables flexible representation of uncertainty \cite{williams2006gaussian}. It has been applied to various power system applications \cite{jalali2022inferring,pareek2020gaussian,mitrovic2023data,10095768,xugp1,pareek2021framework}. One notable drawback, common to all these GP works, is that their applicability has been restricted to small to medium-sized systems. This scale limitation is because exact GP inference has cubic complexity with respect to number of samples \cite{williams2006gaussian}, which grows with system size (or input dimension). Moreover, previous DNN and standard GP models in grid applications do not provide any criteria to apriori decide the number of required training samples. The ability to assess the training quality `on the go' and having a stopping criteria, is of great importance when risk-assessment needs to be done within a short amount of time. 

In this paper, we propose to use a Vertex-Degree Kernel (VDK) based GP model for risk assessment of voltage, given an operating condition and load uncertainty set. First, we establish probabilistic error bounds on  expected value estimation with generic ML models, that applies to probabilistic risk estimation using general GP. Then we show that VDK learns the \textit{voltage-load} function in large-scale systems efficiently by breaking the Kernel function into additive lower-dimensional latent functions based on neighborhoods in the network topology. Further, the VDK-GP model is amenable for Active Learning (AL), an approach to reduce the overall sample complexity of training \cite{settles2009active}, where successive training point containing the \textit{maximum information} about the unknown target function are selected \cite{riis2022bayesian}. AL for standard GP \cite{pareek2021framework} suffers from \textit{curse-of-dimensionality} as the search space grows exponentially with the dimension of input, which is the number of loads. We leverage the additive low-dimensional functions inside VDK-GP's kernel for a novel network-swipe AL algorithm to bypass the curse of dimensionality to further improve its efficiency. Furthermore, we show that VDK-GP's voltage outputs provide a fast alternative to solving standard ACPF for computing the probability of voltage violations, while its theoretical error bound eliminates the need for any  out-of-sample validation. In summary, the main contributions of this study can be delineated as:


\begin{itemize}
\item A conservative probabilistic bound on expected estimation error which establishes VDK-GP's statistical equivalence with ACPF based risk estimation up to the same error threshold. 
We demonstrate that the proposed GP-based model reduces the computational burden in achieving this probabilistic bound.
\item Development of a graph-neighborhood structured kernel, the VDK, for effectively learning the \textit{voltage-load} function in large-scale power systems.  
\item A novel network-swipe AL algorithm for VDK-GP that intelligently selects training data, thereby  eliminating the need for solving numerous ACPF. The proposed AL method also provides a stopping criteria that removes the requirement of out-of-sample testing, thus facilitating its use for operational risk assessment. 
\end{itemize}

To evaluate the proposed method, we conduct benchmark experiments on medium to large-sized power networks, considering uncertain loads at all load buses. Our findings demonstrate that
(a) VDK-GP achieves comparable accuracy with \textit{less than 50\%} of the samples required by standard kernel GP; (b) AL with VDK-GP achieves target model accuracy with \textit{fewer data points} than VDK-GP; and (c) The proposed model provides probabilistic guarantees achieving \textit{ACPF MCS level accuracy while requiring only about $1/15$-th of computational time}.

\section{Related Works}\label{sec:VE}
In power systems, risk assessment seeks to quantify the likelihood and magnitude of operational constraint violations under uncertain injections. For VRA, the objective is to characterize \textit{how nodal voltages $V(\mathbf{s})$ deviate from prescribed limits as the uncertain load vector $\mathbf{s}$ varies over a subspace $\mathcal{S}$ with distribution $\mathcal{P}(\mathbf{s})$.} Formally, one may express a general risk functional as
\begin{align}\label{eq:vra}
\text{VRA} := \mathbb{E}_{\mathbf{s}\sim \mathcal{P}}[\,h(\Delta_V(\mathbf{s}))\,],
\end{align}
where, $\Delta_V(\mathbf{s})$ represents the voltage deviation from limits and $h(\cdot)$ is a violation function (e.g., identity or rectifier).

The challenge in VRA stems from the fact that $V(\mathbf{s})$ is an \emph{implicit, nonlinear mapping} governed by the AC power flow equations. Consequently, even if $\mathcal{P}(\mathbf{s})$ is known, the functional $\text{VRA}$ has no closed-form expression. Existing works therefore rely either on numerical estimates which are time consuming or approximate techniques without guarantees on errors.

Monte Carlo Simulation (MCS) and its variants \cite{allan1981probabilistic,hasan2019existing} remain the most common approach for probabilistic power flow (PPF) and risk estimation. While MCS provides asymptotic convergence, the number of required samples to achieve a desired error bound (e.g., $10^{-3}$ p.u. with $95\%$ confidence) is typically prohibitive for real-time operation—often exceeding $10^5$ to $10^6$ ACPF evaluations \cite{barry2022risk}. Quasi-Monte Carlo and variance-reduction methods \cite{yu2009probabilistic} mitigate but do not eliminate this issue, as their convergence constants depend on the unknown smoothness of $V(\mathbf{s})$.

Alternative numerical methods such as the Point Estimate Method (PEM) \cite{morales2007point,li2015transmission} and cumulant-based or Gaussian-mixture analytical approximations \cite{nijhuis2016gaussian,lin2019fast} attempt to approximate the distribution of voltages. However, these methods lack provable error bounds relative to the true risk functional, and their validity often hinges on restrictive assumptions (e.g., weak correlation among inputs, local linearity of $V(\mathbf{s})$). In particular, PEM-based formulations provide neither consistency nor finite-sample guarantees \cite{li2015transmission}.

\textcolor{black}{An important criticism of the ML methods stems from the fact that ML methods lack performance guarantees. For example, NRLF based MCS will provide bounds on statistical error, which is a function of number of samples used. However, using ML to perform MCS instead of NRLF introduces both approximation error along with statistical error. One way to calculate generalization error of ML methods is described in \cite{AbuMostafa2012Learning}; however, this requires a large number of NRLF solutions for out-of-sample validation-- which in turn defeats the purpose of using ML to speed up the whole process. Ideally, we want \emph{an ML method where total error comprising of approximation error and statistical errors can be calculated, without requiring large number of NRLF solutions.} While interval power flow methods can provide robust performance grantees \cite{ipf}, their NRLF sample requirement grows exponentially with system size since an NRLF solution must be obtained at each corner point of uncertainty hypercube.}

Overall, the literature provides numerous heuristics and approximations for risk evaluation but \textit{almost no framework offering theoretical guarantees}—for example, probabilistic bounds on the estimated voltage risk. This gap motivates the present work, which establishes such guarantees under mild assumptions on load uncertainty and network properties, thereby linking the empirical computation of voltage risk to a provably correct theoretical foundation.

\section{Guarantees on Voltage Risk Assessment}
This section builds on the key idea that voltage risk can be efficiently assessed if a reliable proxy of the NRLF mapping is available. Specifically, we first learn an explicit ML surrogate of the NRLF—one whose approximation error can be theoretically bounded—and then use this learned model to compute the VRA through large-scale numerical evaluations. The proposed GP formulation provides such a proxy:  Once trained, this surrogate allows fast VRA estimation by sampling from the learned model and applying concentration inequalities (e.g., Hoeffding’s) to construct probabilistic bounds on the risk estimate. In essence, we transform the intractable nonlinear risk computation into a two-step procedure—learning a bounded proxy and using it to derive guaranteed probabilistic limits on voltage risk—achieving significant computational savings compared to Monte Carlo–based ACPF simulations (see Fig.~\ref{fig:hoeff}).

\subsection{Voltage Learning Problem}\label{sec:vlearn}
Given a power network with uncertain nodal loads $\mathbf{s} = [p_1, q_1, \dots, p_{|\mathcal{B}|}, q_{|\mathcal{B}|}]$, the voltage learning problem aims to model the functional relationship between load vectors and nodal voltage magnitudes. Let $V_j(\mathbf{s})$ denote the voltage magnitude at node $j$ resulting from the load realization $\mathbf{s}$. This relationship can be expressed as
\begin{align}
    V_j(\mathbf{s}) = f_j(\mathbf{s}) + \varepsilon,
\end{align}
where, $f_j(\cdot)$ represents the deterministic but unknown power flow relation and $\varepsilon$ captures measurement or modeling noise, typically assumed zero-mean with finite variance. The learning task is to approximate the function $f_j(\cdot)$ from a set of training pairs $\{\mathbf{s}^i, V_j^i\}_{i=1}^{N}$ obtained from offline or simulated AC power flow computations.

A range of nonparametric regression models have been applied for this task, with GP regression being a particularly effective choice due to its probabilistic formulation and ability to quantify uncertainty in voltage prediction \cite{williams2006gaussian,liu2022kernel,pareek2021framework}. In a GP-based formulation, the function $f_j(\cdot)$ is modeled as a sample from a stochastic process characterized by a covariance kernel $k(\mathbf{s},\mathbf{s}')$, which encodes the smoothness and locality of voltage responses with respect to load variations. The predictive distribution then provides both the mean voltage estimate and its associated confidence interval for any new load vector\cite{pareek2021framework}. Upon learning, the GP provides mean and variance predictions of the function as
\begin{align}
   \mathbb{E}[f(\mathbf{s})]  & = \mu_f(\mathbf{s}) = \mathbf{k}^T \big [K+\sigma^2_n I \big]^{-1}\mathbf{{V}} \label{eq:vmean} \\
    \mathbb{V}[f(\mathbf{s})] & = \sigma^2_f(\mathbf{s}) = k(\mathbf{s},\mathbf{s})-\mathbf{k}^T[K+\sigma^2_\epsilon I]^{-1}\mathbf{k} \label{eq:vsigma}
\end{align}
Here, $\mathbf{V}$ is the training voltage vector and $K = K(S,S)$ is the estimated kernel matrix over samples of $S$. The vector $\mathbf{k} = [k(\mathbf{s}^1,\mathbf{s}) \dots k(\mathbf{x}^N,\mathbf{x})]$ is obtained by applying the kernel function over $\mathbf{s}$ and $S$\footnote{See appendix in \cite{pareek2021framework} for more details.}. Note that GP not only provides point prediction as $\mu_f(\mathbf{s})$ Eq.~\eqref{eq:vmean}, but also gives confidence level via predictive variance $\sigma^2_f(\mathbf{s})$.

\subsection{Probabilistic Bounds on VRA}\label{sec:guarantee}
As discussed in Section \ref{sec:VE} and Fig. \ref{fig:hoeff}, empirical VRA using MCS of power flow samples is computationally expensive, but can be made faster using GP owing to its closed-form input-output map. We address the question of theoretical guarantee on the performance of GP for VRA. We first derive results for general ML-based methods and then extend these to GP as a special case. Consider a function $h(\cdot)$ defined over node voltage $V$. Let $h_p(\mathbf{s}) = h(V(s))$ indicate its evaluation at voltage $V(s)$ derived from solving NRLF at load $\mathbf{s}$, and let $h_m(\mathbf{s}) = h(\widehat{V}(\mathbf{s}))$ denote its evaluation for voltage derived from a ML predictor (e.g. GP, DNN etc.) at $s$. We then have the following result.
\begin{theorem}\label{thm:1}
 \textbf{Expected Value Estimation Error Bound:} Suppose that a given ML model with output $\widehat{V}$ satisfies
 \begin{align}  \label{eq:model_error_bound}
    \mathbb{P}\big\{\big |V(\mathbf{s}^i) -\widehat{V}(\mathbf{s}^i) \big | > \varepsilon_{m}\big\} \leq \delta 
 \end{align}
  with $\delta\in (0,1)$ for any $\mathbf{s} \in \mathcal{S}$. Let $h(\cdot)$ be a Lipschitz continuous function with Lipschitz constant $L$. Then the error in estimating $h$ with the ML model with $N$ samples is bounded with probability greater than $1-\beta$ as
 \begin{align}
    \big |\mathbb{E}[h_p(\mathbf{s})] - \widehat{\mathbb{E}}[h_m(\mathbf{s})] \big | \leq L \varepsilon_m + 2M \delta + \varepsilon_h 
 \end{align}
 where, $\beta \in (0,1)$, $\varepsilon_h = \sqrt{\frac{4M^2\log{(2/\beta)}}{N}}$, $M$ is a uniform bound on the maximum error that satisfies  $ -M < h(\mathbf{s})< M$, and $N$ is number of samples.
\end{theorem}
Here, $\widehat{\mathbb{E}}[\cdot]$ is the empirical estimate of expectation using $N$ samples. 
The detailed proof is given in the Appendix \ref{app:proofs}. Theorem \ref{thm:1} states that if an ML model is accurate in estimating $V$ (voltage magnitude), then the empirical estimation of expected violation using the ML-generated samples is close to the true expected violation. Further, the use of ML has potential computational speed-up since direct function evaluation is significantly faster than solving traditional ACPF for each sample. Consider the case where $h$ is given by the \texttt{Sigmoid} function to convert deviation from voltage limit $\Delta_V$ calculated using power flow solution as 
\begin{align}\label{eq:sigmoid}
    h_p(\mathbf{s}) = \frac{e^{\Delta_V(\mathbf{s})}}{1+e^{\Delta_V(\mathbf{s})}} - 0.5
\end{align}
Note that \texttt{Sigmoid} function has Lipschitz constant equal to one, and provides information about both extent of violation as well as level of security. We subtract $0.5$ so that $h_p(\mathbf{s}) >0$ implies violation. Theorem \ref{thm:1} provides concentration or error bounds on violation estimation \eqref{eq:vra}. By ensuring preservation of violation level information, the \texttt{Sigmoid} function allows for effective critical load point sampling \cite{lukashevich2023importance}. 

Theorem \ref{thm:1} requires a probabilistic error bound on the ML model given in Eq.~\eqref{eq:model_error_bound}, but validating this bound would require extensive out-of-sample testing, through Hoeffding's inequality\footnote{Minimum number of ACPF-MCS samples $N$ require to obtain statistical error in VRA estimation below $\varepsilon$, with confidence level $\beta$, is given as\\
$N \geq 0.5\log(2/\beta)\varepsilon^{-2}$ \cite{hoeffding1994probability}.}. This poses a challenge when generating ground-truth solutions (e.g., voltage solutions from ACPF) is difficult within the ISO's time-constraints for risk assessment. Contrary to ML models with point-prediction (e.g. DNN), GP automatically offers a measure of confidence around the mean prediction, through the predictive variance in voltage, $\sigma^2_f(\mathbf{s})$ \cite{williams2006gaussian}. This crucial feature enables GP to probabilistically upper-bound voltage solutions ($V(s)$) using predicted mean and variance, 
and eliminates the need for out-of-sample testing. The next corollary extends Theorem \ref{thm:1} for VRA using GP's predictive variance guarantee.

\begin{corollary}\label{coro}
Suppose that the GP assumption holds for PF such that voltage values, for any two arbitrary load vectors, are jointly Gaussian. Then, $\mathbb{P}\{|V(\mathbf{s}) -\widehat{V}(\mathbf{s})|\geq \varepsilon_m \} \leq \delta(\kappa) $ where $\widehat{V}(\mathbf{s}) = \mu_f(\mathbf{s}^i) \pm \kappa\sigma_f(\mathbf{s})$ for any $\varepsilon_m >0$. And with $h(\cdot)$ being \texttt{Sigmoid} function, error in VE using GP is probabilistically bounded as
\begin{align}
   \mathbb{P}\Big \{ \big| \text{VE} - \widehat{\text{VE}} \big| <  \varepsilon_m + 2M \delta(\kappa) + \varepsilon_h \Big \} \geq 1-\beta \nonumber
\end{align}
where, definitions of variables are same as in Theorem \ref{thm:1} and $\delta(\kappa)$ is expected fraction of voltage values outside the range given by $\mu_f(\mathbf{s}^i) \pm \kappa\sigma_f(\mathbf{s})$.
\end{corollary}
The proof follows directly from Theorem \ref{thm:1} and properties of Gaussian distribution. Note that in Corollary \ref{coro}, the GP model error probability $\delta$ is a function of variance multiplier $\kappa$ in $\widehat{V}$. The value of $\delta(\kappa)$ decreases rapidly with increase in $\kappa$ values. At $\kappa =2$ we have $\delta(\kappa) \approx 0.05$ while $\kappa = 4$ will give $\delta(\kappa) \approx 6.334\times 10^{-7}$. As discussed in Fig.~\ref{fig:hoeff} , performing the estimation by solving AC-PF over multiple load samples is not feasible due to high computational burden. Using $N > 820,000$ in Hoeffding's inequality \cite{hoeffding1994probability} and conditions in Corollary \ref{coro}, we will have is $\varepsilon_h = 3\times10^{-3} = \varepsilon_m$\footnote{The confidence bound of the GP is valid for any $\varepsilon_m > 0$. However, for simplicity and to maintain consistency, we chose $\varepsilon_m = \varepsilon_h$.} and confidence of $95\%$ ($\beta =0.05$). Further, using  $\delta(\kappa) = 6.334\times10^{-7}$ and  $M=1$\footnote{As we use \texttt{Sigmoid} function \eqref{eq:sigmoid}, $M=1$ can be used to satisfy the condition $-M < h(s) < M$. More details of $M$ are with proof of Theorem \ref{thm:1} in Appendix \ref{app:proofs}}, the VE will be bounded as $
 \Big |\mathbb{E}[h_p(\mathbf{s})] - \widehat{\mathbb{E}}[h_m(\mathbf{s})] \Big |  \leq 6.001\times10^{-3} $. To generate same accuracy using AC-PF samples,  we will require $205,000$ NRLF solutions that is much more computationally expensive.

Additionally, we can use GP to generate the empirical CDF $\widehat{F}_s$ of violation $h_m(\Delta_V)$. This CDF can provide information on the probability of violation (PoV). formalized below.
\begin{pov}
Given dispatch decision set $\{\mathbf{p}^o_g,\boldsymbol{\alpha}\}$, and node voltage limit $\underline{V}$, PoV is defined as
\begin{align}
 \mathbb{P}_\mathcal{S}\big \{ h(\Delta_V) > 0 \big \}.
\end{align}
\end{pov}
For GP model's applicability, it is important to have confidence that the procedure of performing MCS using GP model instead of NRLF, will not underestimate the PoV. We present the theorem below which certifies that proposed GP-based predictive model will always overestimate the PoV i.e. provide a conservative estimate of security.
\begin{theorem}\label{thm:2}
The GP-based predictive model overestimates probability of voltage violation i.e.$$\mathbb{P}\{h(\mathbf{s}) > 0\} \leq \widehat{\mathbb{P}}\{h_m(\mathbf{s})> 0\}$$ with confidence $1-\beta$.
\end{theorem}
\begin{proof}
    Detailed proof is given in Appendix \ref{app:proofs}
\end{proof}

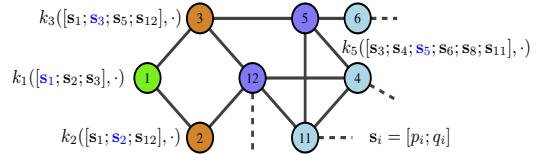
\begin{figure}[t]
\centering
   \vspace{-2em}
\resizebox{\columnwidth}{3.7cm}{
   \begin{tikzpicture}
  \Vertex[IdAsLabel,RGB,color={120,239,30},size = 0.5]{1} 
  \Vertex[IdAsLabel,RGB,color={207,133,40},x=1,y=1,,size = 0.5]{3}
  \Vertex[IdAsLabel,RGB,color={207,133,40},x=1,y=-1,,size = 0.5]{2}
   \Vertex[IdAsLabel,x=2,RGB,color={130,120,246},,size = 0.5]{12}
   \Vertex[IdAsLabel,x=3,y=-1,size = 0.5,,size = 0.5]{11}
   \Vertex[IdAsLabel,x=4,size = 0.5,,size = 0.5]{4}
   \Vertex[IdAsLabel,x=3,y=1,RGB,color={130,120,246},,size = 0.5]{5}
   \Vertex[x=2,y=-1.5,size = 0.5,Pseudo]{7}
   \Vertex[x=4,y=-1,size = 0.2,Pseudo]{13}
   \Vertex[x=5,y=-0.5,size = 0.5,Pseudo]{14} \Vertex[x=5,y=1,size = 0.5,Pseudo]{15}
   \Vertex[x=5,y=2,size = 0.5,Pseudo]{16}
   \Vertex[IdAsLabel,x=4,y=1,size = 0.5]{6}
   \Text[x=-1.5,y=0]{\small $k_1([\textcolor{black}{\mathbf{s}_1};\mathbf{s}_2;\mathbf{s}_3],\cdot)$}
   \Text[x=-0.5,y=-1]{\small  $k_2([\mathbf{s}_1;\textcolor{black}{\mathbf{s}_2};\mathbf{s}_{12}],\cdot)$}
   \Text[x=-0.7,y=1]{\small  $k_3([\mathbf{s}_1;\textcolor{black}{\mathbf{s}_3};\mathbf{s}_{5};\mathbf{s}_{12}],\cdot)$}
   \Text[x=5.5,y=0.5]{\small  $k_5([\mathbf{s}_3;\mathbf{s}_4;\textcolor{black}{\mathbf{s}_{5}};\mathbf{s}_6;\mathbf{s}_{8};\mathbf{s}_{11}],\cdot)$}
   \Text[x=5,y=-1]{\small $\mathbf{s}_i = [p_i;q_i]$}
  \Edge(1)(2)
  \Edge(1)(3)
  \Edge(2)(12)
  \Edge(3)(12)
  \Edge(11)(12)
  \Edge(4)(12)
  \Edge(3)(5)
  \Edge(4)(11)
  \Edge(4)(5)
  \Edge(11)(5)
  \Edge[style={dashed}](12)(7)
  \Edge[style={dashed}](11)(13)
  \Edge[style={dashed}](4)(14)
  \Edge[style={dashed}](6)(15)
  \Edge(5)(6)
\end{tikzpicture}}
\vspace{-1.4em}
    \caption{A part of 118-Bus system showing the idea of vertex degree kernel construction via working over loads of a given nodes and its immediate neighbors. In Eq.~\eqref{eq:VDK}, we use $\mathbf{x}_b$ to represents all the load variables in $b$-th sub-kernel as $k_1(\mathbf{x}_1,\cdot)= k_1(\mathbf{s}_1,\mathbf{s}_2,\mathbf{s}_3,\cdot)$. For brevity, second input element in kernel function is represented using $(\cdot)$.}
    \vspace{-1.5em}
    \label{fig:VDK}
\end{figure}

\section{Vertex-Degree Kernel and Active Learning}
Despite their flexibility, standard kernel choices (e.g., squared exponential) often require a large number of training samples to achieve accurate voltage predictions, as kernel hyperparameter optimization scales poorly with system dimension. This motivates the development of structure-aware or physics-informed kernels that exploit the topology and locality properties of the power network to improve sample efficiency without sacrificing predictive fidelity \cite{liu2022kernel,xugp1,xugp2}.

We design an additive GP kernel using sub-kernels, each being a squared exponential-kernel limited to the neighborhood of a nodal load, i.e., for node $j$ we have set $\mathbf{x}_j=\{s_i|i = j \text{~or~} (ij)\in \mathcal{E}\}$. The intuition behind the proposed kernel design is that a node's voltage is affected by all loads but their \emph{correlated effect}\footnote{`correlated' effect is the change in voltage due to simultaneous change in two or more nodal loads.} is limited to loads that are near one another. In other words, effect of two far-away loads on nodal voltages can be considered uncorelated/independent. As maximum degree of a node is much less than the size of a power grid, each sub-kernel is low-dimensional. The complete additive kernel, the sum of these sub-kernels, is termed as VDK, and is defined as
\begin{align}\label{eq:VDK}
    k_{v}(\mathbf{s}^i,\mathbf{s}^j) =  \sum^{|\mathcal{B}|}_{b =1} k_b(\mathbf{x}^i_b,\mathbf{x}^j_b) 
\end{align}
Here, $k_b(\mathbf{x}^i_b,\mathbf{x}^j_b)$ is the sub-kernel working over $\mathbf{x}_b \subset \mathbf{s}$. Note that by relying on the grid structure, neighborhood based correlations are kept intact in VDK, but complex design choices for kernels are avoided. Fig. \ref{fig:VDK} shows the idea of VDK construction. Each sub-kernel $k_b(\cdot,\cdot)$ has hyper-parameters $\boldsymbol{\rm \theta}_b$ that form the full hyper-parameter vector for VDK as $\boldsymbol{\theta} = [\boldsymbol{\rm \theta}_1, \cdots, \boldsymbol{\rm \theta}_{|\mathcal{B}|}]$.  As the sum of valid kernel functions is a kernel function, standard exact inference can be performed via optimizing MLL using $k_{v}(\cdot,\cdot)$ \cite{lu2022additive,duvenaud2011additive,williams2006gaussian}. 
However, as \textit{square exponential kernel} has two hyper-parameters \eqref{eq:se_kernel}, the total number of hyper-parameters in VDK \eqref{eq:VDK} will be twice the number of network nodes. In the next section, we show that the neighborhood based additive form of VDK lends itself to a simple active learning algorithm for fast hyperparameter estimation.

\color{black}
\begin{remark}\textbf{Uncertainty and Correlation Handling:}
    It is crucial to note that the uncertainty quantification provided by the GP framework is inherently agnostic to the underlying probability distribution of the input variables. Once trained, the GP posterior provides a deterministic, closed-form mapping for the predictive mean $\mu_f(\mathbf{s})$ and variance $\sigma_f^2(\mathbf{s})$ for any arbitrary net injection vector $\mathbf{s}$. Consequently, uncertainty propagation is executed entirely at the sampling level. Evaluating voltages and associated risk metrics simply requires drawing point-wise samples from the target input distribution and evaluating them using the GP's closed-form expressions. Because this evaluation process does not assume statistical independence among the inputs, highly correlated joint distribution $\mathcal{P}(\mathbf{s})$ (renewable generation and load uncertainties) can be directly sampled and passed through the model. The VDK complements this sampling capability by explicitly encoding physical locality and topological dependencies into the covariance structure $k(\mathbf{s}, \mathbf{s}^{\prime})$.
\end{remark}

\color{black}

\subsection{Active Learning for VDK-GP}
We are interested in rapid estimation of $f(\mathbf{s})$ ($j$-th node's voltage function) by learning the hyper-parameter $\boldsymbol{\theta}$, and further in determining a tractable stopping criterion for terminating the learning that does not rely on computationally expensive out of sample AC-PF solutions. 

To answer the first part, we propose an active learning (AL) \cite{settles2009active} mechanism that sequentially selects training samples to maximize the information gained about the voltage function, using predictive variance as an indicator to terminate the learning process. In GP modeling (or Bayesian setting in general) `information gain' is used as a measure to quantify the informativeness of a training set or sample. Let, $\mathcal{A} \subset \mathcal{L}$ where $\mathcal{L}$ is the complete training set or space. The Information gained by samples in $\mathcal{A}$ is the information theoretic mutual information between voltage function $f$ and $\mathbf{V}_\mathcal{A}$ (vector of voltage samples) \cite{srinivas2012information}, and is given by $I(\mathbf{V}_\mathcal{A};f) = (1/2)\log|I + \sigma^{-2}_\epsilon K_\mathcal{A}|$. Here $K_\mathcal{A} $ is the kernel matrix constructed using samples in set $\mathcal{A}$. Importantly, finding the best $\mathcal{A}$ with given cardinality $|\mathcal{A}| <|\mathcal{L}|$ is an \textit{NP-hard} problem \cite{srinivas2012information}. However, two results facilitate tractable AL. First, in \cite{krause2012near} information gain has been shown to be a  \textit{submodular} function of set $\mathcal{A}$, implying that greedy algorithm for sample selection is at least within $\sim 63\%$ of optimal solution. Second, the information gain in a new sample $s$ is given  by predictive variances Eq.~\eqref{eq:vsigma}. Hence, the  next training sample, for $j$-th node voltage function learning, can be obtained by solving 
\begin{align}\label{eq:al}
    \mathbf{s}^{t+1}= \argmax_{\mathbf{s} \in \mathcal{L}} \,[{\sigma^t_f}(\mathbf{s})]^2
\end{align}
Here, $\sigma^t_f(\cdot)$ is predictive standard deviation of the GP trained on first $t$ samples, as given by Eq.~\eqref{eq:vsigma}. As eluded in the introduction, for large networks, the non-convex function makes Eq.~\eqref{eq:al} quickly intractable for standard GP \cite{pareek2021framework}, as the input vector in Kernels has size ${2|\mathcal{B}|}$ \cite{frazier2018tutorial}. While VDK is separable in terms of lower-dimensional sub-kernels $k_b$ given in Eq.~\eqref{eq:VDK}, they have overlapping input groups for any two sub-kernels $k_{b_i}, k_{b_j}$ at nodes $i,j$ that are within two graph hops. This means that a simple parallelization into sub-kernels isn't possible for AL. 
\begin{figure}[t]
\centering
   \begin{tikzpicture}
  \Vertex[IdAsLabel,size = 0.6,y=2.5,x=-1.2]{1} 
  \Vertex[IdAsLabel,RGB,color={120,239,30},x=-0.5,y=1,size = 0.6]{3}
  \Vertex[IdAsLabel,RGB,color={120,239,30},x=-2,y=1,,size = 0.6]{2}
   \Vertex[IdAsLabel,x=-2,y=-0.3,RGB,color={130,120,246},,size = 0.6]{12}
   \Vertex[IdAsLabel,x=-0.5,y=-0.3,RGB,color={130,120,246},,size = 0.6]{5}
   \Vertex[x=1.3,y=2.0,size = 0.05,Pseudo]{6}
   \Vertex[x=4.,y=1.2,size = 0.05,Pseudo]{7}
    \Vertex[x=1.0,y=0.6,size = 0.05,Pseudo]{8}
   \Vertex[x=4.8,y=-0.3,size = 0.05,Pseudo]{9}
    \Text[x=3,y=2.2]{\small ${\widehat{\mathbf{x}}^{t+1}_{\mathcal{D}_0}} = \argmax\limits_{\mathbf{s}_1} \sigma^t_f \big (\widehat{\mathbf{x}}_{\mathcal{D}_0}\textcolor{black}{\widehat{\mathbf{x}}^t_{\mathcal{D}_{1\dots d}}} \big ) $}
    \Text[x=3,y=0.8]{\small ${\widehat{\mathbf{x}}^{t+1}_{\mathcal{D}_1}}  = \argmax\limits_{\mathbf{s}_2,\mathbf{s}_{3}} \sigma^t_f \big (\textcolor{black}{\widehat{\mathbf{x}}^{t+1}_{\mathcal{D}_0}},\widehat{\mathbf{x}}_{\mathcal{D}_1}, \textcolor{black}{ \widehat{\mathbf{x}}^t_{\mathcal{D}_{2\dots d}}} \big )$}
    \Text[x=1,y=-0.4]{\small $\cdots \cdots \cdots $}
    \Text[x=4.5,y=-0.4]{\small $\cdots \cdots \cdots $}
  \Edge(1)(2)
  \Edge(1)(3)
  \Edge(2)(12)
  \Edge(3)(12)
  \Edge(3)(5)
\Edge[color=green!70!black,opacity=.9,lw=1pt,Direct,bend=8](6)(7)
\Edge[color=green!70!black,opacity=.9,lw=1pt,Direct,bend=8](8)(9)
\end{tikzpicture}
    \caption{Description of steps within $(t+1)^\text{th}$ iteration of network-swipe Algorithm \ref{alg:gp_al} for AL. At each step, only loads at a fixed graph distance from $j$ are treated as variables (colored in black) for information maximization, while others are kept fixed (colored in black). For example, $\overline{\mathbf{x}}_{\mathcal{D}_1} = \{\mathbf{s}_2, \mathbf{s}_3\}$. Once those loads are optimized, their values are updated before the next step. In second step, updated output of the first step  $\widehat{\mathbf{x}}^{t+1}_{\mathcal{D}_0}$ are used along with values from previous iteration at all other loads $\widehat{\mathbf{x}}^t_{\mathcal{D}_{2\dots d}}$, as given in Eq.~\eqref{eq:network_swipe}.}
    \vspace{-1.5em}
    \label{fig:newtork-swipe}
\end{figure}
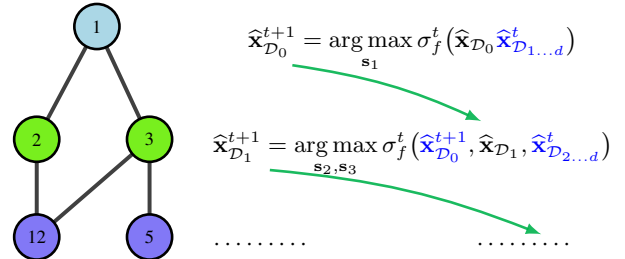
Instead, we propose a block-descent type iterative network-swipe algorithm to exploit VDK's form for optimizing Eq.~\eqref{eq:al}. 
At each iteration $t$ of the network-swipe algorithm, at the first step, we solve Eq.~\eqref{eq:al} with respect to $\mathbf{x}^t_j$ (load at the node where voltage is being predicted), while keeping all other loads fixed. The load at $j$ is updated with the optimized value. In the second step, we solve Eq.~\eqref{eq:al} for all loads at nodes $i$ such that $(ij)\in \mathcal{E}$ (1 hop neighbors of $j$). All other loads are kept unchanged. In the next step, loads at two-hop neighbors of $j$ are chosen to solve Eq.~\eqref{eq:al}, and so on till all loads have been updated. For elucidation, we use $\mathcal{D}^j_r, 0\leq r\leq d$ to denote distinct node groups at a graph distance of $r$ from node $j$, with max-distance $d$. Hence, $\mathcal{D}^j_0 = \{j\}$, while $\mathcal{D}^j_r = \{i: (\mathcal{D}^j_{r-1},j)\in\mathcal{E}\}$. 
Mathematically, the $t^\text{th}$ iteration of network-swipe solves the following non-linear optimization problems sequentially for $i = 0,1 \dots d $ 
\begin{align}\label{eq:network_swipe}
    \widehat{\mathbf{x}}^{t+1}_{\mathcal{D}_i} & = \argmax_{\widehat{\mathbf{x}}_{\mathcal{D}_i} \in \mathcal{L}_i} \, \sigma^t_f (\widehat{\mathbf{x}}^{t+1}_{\mathcal{D}_{0 \dots i-1}},\widehat{\mathbf{x}}_{\mathcal{D}_i},\widehat{\mathbf{x}}^t_{\mathcal{D}_{i+1 \dots d}}) 
\end{align}
Here, $\widehat{\mathbf{x}}_{\mathcal{D}_i} = \{\mathbf{s}_j|j \in \mathcal{D}_i\}$. Also, $\mathcal{L}_i$ represents the hyper-cube slice with respect to loads present in $\widehat{\mathbf{x}}_{\mathcal{D}_i}$. The algorithm then starts a new iteration (${t+1}^\text{th}$) to determine the next sample. The pseudo-code for AL is listed in Algorithm \ref{alg:gp_al} and a graphical representation of steps for target node $j$ is shown in Fig. \ref{fig:newtork-swipe}. 
\begin{algorithm}[b]
\caption{Network-Swipe Algorithm for AL}
\label{alg:gp_al}
\begin{algorithmic}[1]
\Require $T$, $\mathcal{D}$, $\{\mathbf{s}^1,V^1\}$, $\sigma^2_{threshold}$, $T_{max}$
\State Initialize GP model with VDK \eqref{eq:VDK}
  \While{$\sigma^2_f(\mathbf{s}^t) \geq \sigma^2_{threshold}$}
  \vspace{0.3em}
  \State \text{Solve} \eqref{eq:network_swipe} for $\widehat{\mathbf{x}}^{t+1}_{\mathcal{D}_i}$, sequentially for $i = 1\dots d$ 
  \State Solve ACPF for load $\mathbf{s}^{t+1}$ to get $V^{t+1}$ 
  \State Update GP model with $(\mathbf{s}^{t+1}, V^{t+1})$; $t = t+1$
  \State \textbf{If} \texttt{runtime} $> T_{max} $ \textbf{then} \texttt{break}
\EndWhile
 \Output Compute $\mu_f(\mathbf{s}), \sigma_f^2(\mathbf{s})$ for final GP 
\end{algorithmic}
\end{algorithm}

In Algorithm \ref{alg:gp_al}, we have used a time budget of $T_{max}$, and a predictive variance threshold of $\sigma^2_{threshold}$. While we present a single sequence of network swipe steps to determine the next injection sample $s^t$, multiple swipes can be performed across network to improve the information gain in $s^t$. Further, for solving Eq.~\eqref{eq:network_swipe}, in Algorithm \ref{alg:gp_al} we use function evaluation with different batch sizes and select the best candidate of injection among the random samples. 
Further, the function evaluation-based approach will allow to build parallelizable optimization methods, helping to scale and improve performance of the proposed network-swipe AL in future works. In the next section, we provide guarantees on using VDK-GP for risk estimation in the grid.  

It is crucial to emphasize the effectiveness of the proposed network-swipe AL method, which is also related to the incorporation of the VDK structure. This Algorithm \ref{alg:gp_al} enables the optimization problem, as defined in \eqref{eq:network_swipe}, to remain low-dimensional. Utilizing a conventional GP kernel, standard active learning methods typically encounter the curse of dimensionality during sample-based optimization, particularly when seeking maximum variance across a load space of dimension $2|\mathcal{B}|$, where $|\mathcal{B}|$ denotes the number of buses within the system. Consequently, the proposed network-swipe AL design requires sample-based optimization over load variables set at a particular depth (defined as $\mathcal{D}^j_r$) at a time. This low-dimensionality of optimization is particularly advantageous for applying proposed active learning method to large-scale networks.

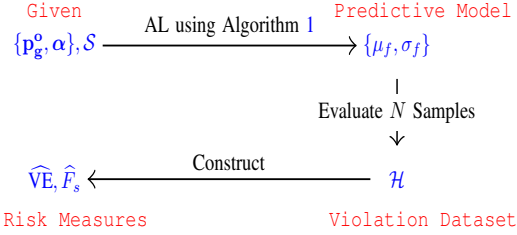
\begin{figure}
    \centering
\resizebox{0.8\columnwidth}{3.2cm}{
   \begin{tikzpicture}
   \node [rectangle, minimum width=1cm, minimum height=1cm] (a) {\textcolor{black}{$\{\mathbf{p^o_g},\boldsymbol{\alpha}\}, \mathcal{S}$}};
   \node [above =-2mm of a] (at) {\textcolor{red}{\texttt{Given}}};
   \node [right = 1.9in of a,rectangle, minimum width=1cm, minimum height=1cm] (b) {\textcolor{black}{$\{\mu_f,\sigma_f\}$}};
   \node [above =-2mm of b,xshift=5mm] (bt) {\textcolor{red}{\texttt{Predictive Model}}};
   \node [below = 0.4in of b,rectangle, minimum width=1cm, minimum height=1cm] (c) {\textcolor{black}{$ \mathcal{H}$}};
   \node [below = -1.5mm of c,xshift=5mm] (ct) {\textcolor{red}{\texttt{Violation Dataset}}};
   \node [below = 0.4in of a,rectangle, minimum width=1cm, minimum height=1cm] (d) {\textcolor{black}{$\widehat{\text{VE}},\widehat{F}_s$}};
   \node [below = -1.5mm of d,xshift=4mm] (dt) {\textcolor{red}{\texttt{Risk Measures}}};
    \draw [-{Classical TikZ Rightarrow[length=1.5mm]},thick] (a) -- node [midway,above]{\textcolor{black}{AL using Algorithm \ref{alg:gp_al}}} (b);
    \draw [-{Classical TikZ Rightarrow[length=1.5mm]},thick] (b) -- node [midway,fill=white]{\textcolor{black}{Evaluate $N$ Samples}}(c);
    \draw [-{Classical TikZ Rightarrow[length=1.5mm]},thick,] (c) -- node [midway,above] {\textcolor{black}{Construct}}  (d);
   \end{tikzpicture}} 
   \vspace{-0.5em}
    \caption{Flowchart of risk measure calculation using AL-VDK based voltage learning. Here, $\mathcal{H}_ = \{h_m(\Delta^i_V)\}_{i=1}^{N}$ and $\widehat{F}_s$ is the cumulative density function (CDF) of violation $h_m(\Delta_V)$ random variable. This is then used directly to estimate probability of voltage violation $\widehat{\mathbb{P}}\{h_m(\mathbf{s})> 0\}$.} 
    \label{fig:overall}
\vspace{-1em}
\end{figure}

\section{Results and Discussion}
In this section, simulation results demonstrate a) that our Graph-structured kernel (VDK)  \eqref{eq:VDK} outperforms standard GP model \cite{pareek2021framework} in voltage prediction at same sample complexity, b) AL with VDK (\textbf{AL-VDK}) efficiently learns voltage functions with acceptable error using fewer samples than VDK and c)  AL-VDK voltage predictor exhibits significantly lower time complexity than NRLF for statistical estimation of voltage violation (VE) (Corollary \ref{coro}), while the proposed model is a conservative estimator of risk of violation (Theorem \ref{thm:2}). The VE value indicate the extent by which voltage goes beyond the lower voltage limits, calculated using the \texttt{Sigmoid} \eqref{eq:sigmoid} function. To obtain the node voltages for given uncertainty set $\mathcal{S}$ and decision set $\{\mathbf{p}^o_g,\boldsymbol{\alpha}\}$, we use
\texttt{PowerModels.jl}, for running ACPF. For this, upon sampling a load vector $\mathbf{s}\in \mathcal{S}$, we update the generator points in the data file using the participation factors i.e. $\mathbf{p}_g = \mathbf{p}^o_g + \boldsymbol{\alpha}\Vec{\Delta \mathbf{s}}$ with $\Vec{\Delta \mathbf{s}}$ being sum of load change from base-point. To validate the model, 1000 out-of-sample testing points are used in all cases unless stated otherwise. Three different systems from \texttt{pglib-library}\cite{pglib} are used for validation (118-Bus, 500-Bus, and 1354-Bus). We use the \texttt{Square Exponential Kernel}, both for standard GP and VDK-GP and model them in $\texttt{Julia}$. Additionally, we use a \textbf{DNN}, deep neural network, of three layers and 1000 neurons in each layer \cite{gao2023physics}\footnote{We use \texttt{Flux.jl} with standard settings of \texttt{ADAM} solver to optimize hyper-parameters. Batch size is 5 and Epochs are set as 200.} for comparison. We use mean absolute error (MAE) to validate the performance of proposed models:  $\text{MAE}  = M^{-1} \sum_{i} \big | \mu_f(\mathbf{s}^i) - V(\mathbf{s}^i) \big |$ for $i = 1 \dots M$.

\begin{figure}[t]
    \centering
  \includegraphics[width=\columnwidth]{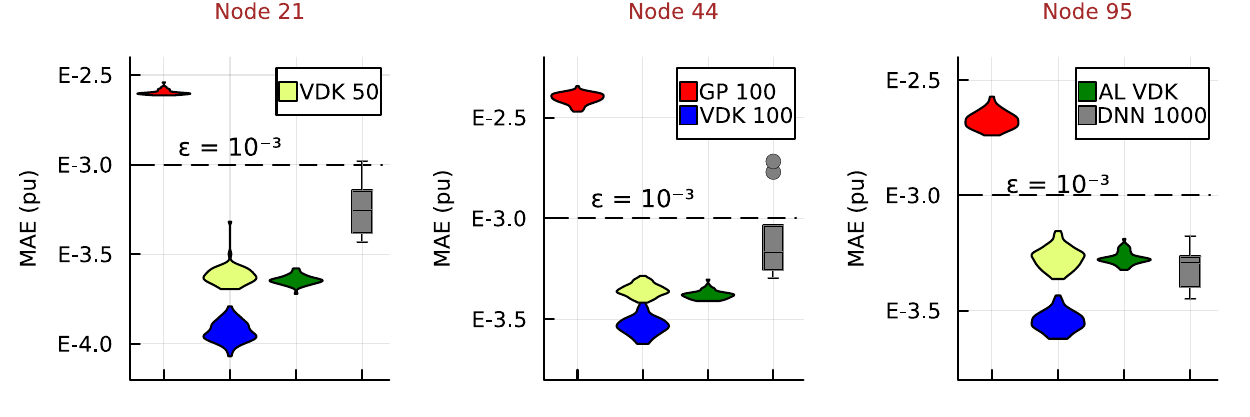}
    \vspace{-2em}
    \caption{Comparison of MAE performance of different methods, demonstrating the efficiency and low sample-complexity of AL-VDK, on three different nodes of 118-Bus system. GP, VDK and AL-VDK results are of 50 trials, and DNN results are of 10 trials. AL-VDK uses significantly fewer samples as dictated by Algorithm \ref{alg:gp_al}. AL-VDK training samples for all 50 trials are within 43 -- 48, 43 -- 47 and 42 -- 47 for nodes 21, 44, and 95 respectively.}
    \label{fig:MAE_compare_118}
    \vspace{-1em}
\end{figure}

\subsubsection*{\textbf{118-Bus System}}
The load space is constructed by varying individual real power loads within $\pm$10\% and load power factors between 0.9 times the base power factor and 1.0. Fig. \ref{fig:MAE_compare_118} compares the MAE performance of four different methods: GP, VDK, DNN, and AL-VDK. First, GP fails to achieve MAE below the desired threshold of $1E-3$ using 100 training samples. Further as shown, AL-VDK achieves competitive MAE performance while using significantly fewer training samples than the other methods. Notably, AL-VDK's sample usage is dynamically determined by Algorithm \ref{alg:gp_al}, leading to improved data efficiency. In contrast, DNNs require a larger set of 1,000 samples for each trial to achieve an error of the same order, highlighting their high training data requirement. Also, note that AL-VDK performs better than VDK with 50 samples, showing AL's capability to reduce MAE efficiently with an increase in training samples. The average time taken by a complete trial of the AL-VDK algorithm is 5.8 seconds (without testing) compared to 8.3 seconds for 100-sample VDK. These results demonstrate that AL-VDK offers a promising combination of accuracy and efficiency, making it a valuable tool for scenarios where limited ACPF data is available or acquiring PF data by running NRLFs is time-consuming.

\begin{table}[t]
    \centering
     \caption{Training Sample Requirement and Risk Estimation Results in 500-Bus System}
    \begin{tabular}{c|ccc}
        & Training Samples & Time(s)  & $\Delta\text{VE}\times 10^{-4}$  \\
        \hline
     4  & 67 - 70 & 28 - 30     &  7.8 $\pm$ 0.5 \\
    181  &71 - 76 & 30 - 33    & 8.0 $\pm$ 0.2  \\
     268  & 102 - 109 & 53 - 58    & 7.9 $\pm$ 0.2 \\
      320  & 72 - 76 & 30 - 33    & 7.8 $\pm$ 0.4 \\
          321  & 70 - 77 & 30 - 33  & 6.8 $\pm$ 0.5 \\
          \hline
     \multicolumn{4}{c}{Mean evaluation time for 82000 samples is $\approx$ 33.2 sec}\\
      \multicolumn{4}{c}{NRLF running time for 20500 samples is $\approx$ 4205 sec}\\
     \hline 
         \multicolumn{4}{l}{$\Delta\text{VE}$: Difference in VE values using NRLF and AL-VDK}\\
    \end{tabular}
        \vspace{-1.5em}
    \label{tab:500}
\end{table}

\begin{table}[t]
    \centering
    \caption{Estimated Value Probability of Violation (POV) for 500-Bus System Voltages with Difference in Estimation using AL-VDK}
    \begin{tabular}{c|ccccc}
      & \multicolumn{5}{c}{\textbf{Node}} \\
         \cline{2-6}
      \textbf{PoV}   & 4 & 181 & 268 & 320 & 321 \\
         \hline
       Estimated$^\dagger$  &  0.0487 & 0.0115 & 0.6879 & 0.8092 & 0.9108  \\
       Difference$^\#$ &  $+$0.04 & $+$0.01 &  $+$0.16 &  $+$0.15 &  $+$0.10\\
       \hline
       \multicolumn{6}{l}{$^\dagger$using 82000 and mean over AL trials;}\\
       \multicolumn{6}{l}{$^\#$between 20050 ACPF and  82000 AL-VDK evaluations}\\
        \multicolumn{6}{l}{$^\#$Positive Difference $\implies$ Overestimation}
    \end{tabular}
    \vspace{-2em}
    \label{tab:500prob}
\end{table}

\subsubsection*{\textbf{500-Bus System}}
\begin{figure*}[t]
    \centering
    \includegraphics[width=\textwidth]{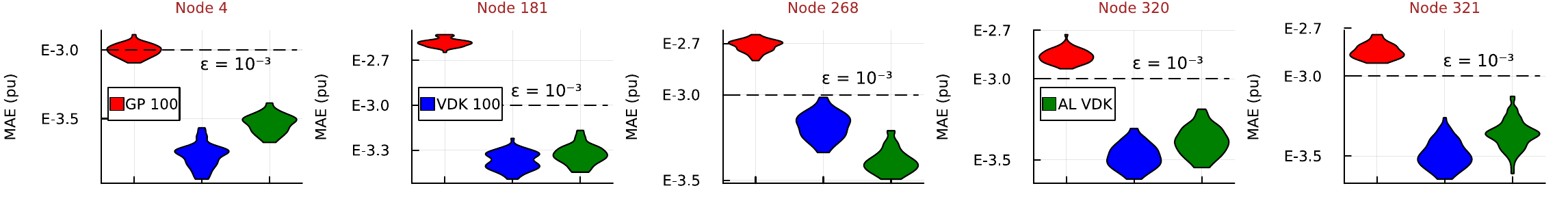}
    \vspace{-2em}
    \caption{Comparison of MAE performance of different methods, demonstrating the efficiency and low sample-complexity of AL-VDK, on three different nodes of 500-Bus system. AL-VDK uses significantly fewer samples as dictated by Algorithm \ref{alg:gp_al} and details are given in Table \ref{tab:500}. Our target is to achieve MAE $< 10^{-3}$ (0.1\% error for a 1000kV system), and ACPF samples are generated till that threshold is reached.}
    \label{fig:MAE_compare_500}
    \vspace{-1.5em}
\end{figure*}

The load space is constructed by varying individual real power load within $\pm10\%$ and load power factor between 0.95 times of base power factor and 1.0. Besides the MAE accuracy results, in this subsection we present the risk estimation results on five different node voltages. These node indices are selected as the voltages at these nodes have the  highest variance among 20500 ACPF solutions. Fig. \ref{fig:MAE_compare_500} compares the MAE performance of different methods GP, VDK, and AL-VDK while Table \ref{tab:500} shows details of AL-VDK result for all five nodes. Firstly, the MAE performance of AL-VDK  is comparable to that of VDK even when it uses $\sim$ 70 training samples instead of 100 used by VDK for most nodes in Fig. \ref{fig:MAE_compare_500}. Further, using similar number of samples, AL-VDK outperforms VDK for node 268 which has higher voltage variance. This show that at same sample complexity, the worst performing trial of AL-VDK is better than mean performance of VDK. 

To calculate risk levels and $\Delta \texttt{VE}$ presented in Table \ref{tab:500},  we fix the target that error in violation estimation must be less than $6 \times 10^{-2}$. The estimation error here means the difference between true VE and proposed ML-based VE i.e. $\Big |\mathbb{E}[h_p(\mathbf{s})] - \widehat{\mathbb{E}}[h_m(\mathbf{s})] \Big |$, from  Theorem \ref{thm:1}. To estimate violation with statistical error less than $6 \times 10^{-2}$, at least 20500 NRLF solutions\footnote{This is because $\varepsilon = 6\times10^{-3}$ will require more than 200000 NRLF solutions which is computationally prohibitive. Also, note that from Corollary \ref{coro}, to achieve $\Big |\mathbb{E}[h_p(\mathbf{s})] - \widehat{\mathbb{E}}[h_m(\mathbf{s})] \Big | \leq 6 \times 10^{-2}$, $\varepsilon_h$ should be set at $3\times10^{-2}$ and thus we require 82,000 AL-VDK evaluations.} are required as per Hoeffding's inequality and 82000 AL-VDK model evaluation are needed as per Corollary \ref{coro}. The 20500 NRLF trials took more than one hour (4205 sec.) on a Mac M1 Max with 32GB RAM. On same machine it takes only $\sim$33.2 seconds to obtain 82000 voltage predictions for a node. Considering the training time and evaluation time, it takes less than 90 seconds to estimate risk levels with error $\Delta\texttt{VE}$ of the order $10^{-4}$, which is two order of magnitude lower than acceptable statistical error level of $6\times 10^{-2}$. Note that, $\Delta\texttt{VE}$ represents the difference between NRLF-based statistical estimation using 20500 samples and AL-VDK based statistical estimation using 82000 samples. The two order of magnitude difference in $\Delta\texttt{VE}$ and acceptable $\Big |\mathbb{E}[h_p(\mathbf{s})] - \widehat{\mathbb{E}}[h_m(\mathbf{s})] \Big |$ signifies that AL-VDK induced error in estimation will not affect the quality of the statistical estimation. Thus, NRLF samples can be replaced by AL-VDK evaluations as shown in Fig. \ref{fig:overall}. Also note that proposed method achieves $\Delta \texttt{VE}$ and MAE of same order.

\begin{figure}[t]
    \centering
    \includegraphics[width=0.8\columnwidth]{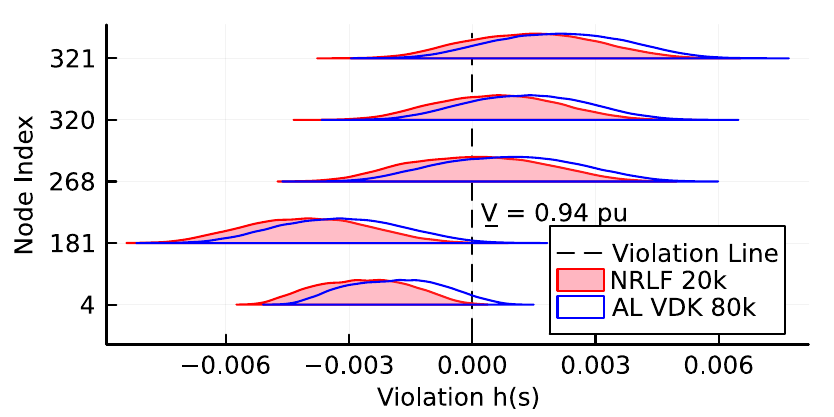}
    \vspace{-1.0em}
    \caption{Distributions of violation $h(\mathbf{s})$ obtained using 20050 NRLF solutions and 82000 AL-VDK evaluations after a using a random training instance. The right hand side shift of black distributions shows that proposed AL-VDK always provides an overestimation of risk, as proven in Theorem \ref{thm:2}.}
    \vspace{-1.5em}
    \label{fig:violations}
\end{figure}

As discussed in Section \ref{sec:guarantee}, we would want an ML model to be conservative in risk estimation i.e. estimated probability of violation must be more using an ML model than estimated using NRLF solution. Fig. \ref{fig:violations} shows the violation distributions and violation probability values calculated using NRLF and AL-VDK models. As black curves are shifted towards right for all nodes, it is clear that AL-VDK overestimates the risk compares to NRLF as proved in Theorem \ref{thm:2}. Similarly, Table \ref{tab:500prob} shows that difference between AL-VDK based estimation and ACPF based estimation of probability of violation is always positive. Also, results show that voltage at node 4 and node 181 are probabilistically secure, if maximum acceptable violation is considered to be 5\%. 
 
\subsubsection*{\textbf{1354-Bus System}} The construction of the load space is similar to that of the 500-Bus case. Table \ref{tab:1354} presents the details of AL-VDK results across 10 trials on two distinct nodes. We consider a lower limit of 0.90 pu, and the node index is chosen based on a variation exceeding 0.010 pu (minimum - maximum). To manage the sample requirement, we set $\varepsilon=10^{-1}$ due to the substantial time needed for solving ACPFs in the 1354-Bus system. In Table \ref{tab:1354}, both nodes exhibit $\widehat{\text{VE}} > 0$, indicating conservativeness of the AL-VDK based estimates. Additionally, we achieve a VE difference of $10^{-4}$, mirroring the outcomes in the 500-Bus system. The total time taken for learning and VE estimation is less than 200 seconds for a node, a significant reduction compared to the approximately one hour needed for solving 2050 PF solutions. Consequently, the proposed method attains a 20-fold reduction in the time complexity for obtaining VE with an error of order $10^{-4}$. These results underscore the applicability of the proposed method, even for  large-scale systems.

\begin{table}[t]
    \centering
     \caption{Risk Estimation Results in 1354-Bus System}
    \begin{tabular}{c|ccc}
        & Samples & Time(s) & $\Delta\text{VE}\times 10^{-4}$  \\
        \hline
     183  & 77 - 81 & 159 - 168   &  8.0 $\pm$ 0.5 \\
    287  & 77 - 81 & 154 - 164  & 8.2 $\pm$ 0.2 \\
          \hline
      \multicolumn{4}{c}{Mean evaluation time for 8200 samples is $\approx$ 29.8 sec}\\
     \multicolumn{4}{c}{NRLF running time for 2050 samples is $\approx$ 3879 sec}\\
     \hline 
         \multicolumn{4}{l}{$\Delta\text{VE}$: Difference in VE using NRLF and AL-VDK}\\
    \end{tabular}
        \vspace{-1.5em}
    \label{tab:1354}
\end{table}

\section{Conclusions and Future Work}
This work developed a provably reliable and computationally efficient framework for probabilistic voltage risk estimation using GP learning. By deriving explicit error and confidence bounds on voltage risk metrics, we bridged the methodological gap between MCS–based uncertainty quantification and modern learning-based surrogates. The VDK, grounded in network topology, enabled tractable GP inference by decomposing voltage–load relationships into local subspaces, while the proposed network-swipe active learning algorithm ensured sample efficiency and eliminated the need for out-of-sample validation. Together, these advances yield a conservative risk estimator that preserves operational security—guaranteeing that the predicted probability of violation does not underestimate the true risk. Quantitatively, the approach achieved mean absolute voltage errors below $10^{-3}$ p.u. across standard test systems and reproduced Monte Carlo–level voltage violation statistics using 15$\times$ fewer ACPF computations and 120× lower evaluation time. The results validate that probabilistic guarantees and scalable learning can coexist, offering an interpretable alternative to opaque deep networks for safety-critical power system operation.

Future work will focus on extending the probabilistic guarantee framework to non-Gaussian uncertainty models and time-coupled stochastic processes in renewable-rich grids. Further, integrating VDK-GP within optimization and planning loops can enable real-time decision-making under quantified risk, while distributed and sparse GP formulations may further reduce complexity for national-scale power systems. Ultimately, this line of work moves toward learning-based yet certifiably reliable power system analytics, where uncertainty quantification is both fast and trustworthy.

\bibliographystyle{IEEEtran}
\bibliography{main}

\appendices

\section{Mathematical Preliminaries and Notation}\label{app:notation}

Consider a power grid $\mathcal{G} = (\mathcal{B}, \mathcal{E})$, where $\mathcal{B}$ denotes the set of buses and $\mathcal{E}$ denotes the set of transmission lines. Each line $(ij) \in \mathcal{E}$ is characterized by an impedance $R_{ij} + \hat{j}X_{ij}$. The complex node voltage is denoted as $v_j = V_j \angle \theta_j$, with magnitude $V_j$ and phase angle $\theta_j$. 

We adopt the convention that net load at node $j$ is defined as $\text{load} = \text{demand} - \text{generation}$, and denote the complex power injection as $s_j = [p_j; q_j]$, where $p_j$ and $q_j$ represent the real and reactive components, respectively. The set of all admissible load vectors forms the uncertainty subspace $\mathcal{S}$.

The steady-state AC power flow equations express the implicit mapping between nodal injections $\mathbf{s}$ and voltages $\mathbf{v}$:
\begin{align}\label{eq:pf_appendix}
\forall j \in \mathcal{B}, \quad
s_j = \sum_{i:(ij)\in\mathcal{E}} v_j \frac{(v_j^* - v_i^*)}{R_{ij} + \hat{j}X_{ij}}.
\end{align}

Given this implicit relation, node voltages $V_j(\mathbf{s})$ depend nonlinearly on the stochastic load vector $\mathbf{s}\sim\mathcal{P}(\mathbf{s})$. Voltage risk assessment (VRA) seeks to quantify the expected or probabilistic violation of voltage limits under this uncertainty. Defining voltage deviation $\Delta_V = \underline{V} - V(\mathbf{s})$ (for lower limits) or $V(\mathbf{s}) - \overline{V}$ (for upper limits), the general VRA functional is written as \eqref{eq:vra} where $h(\cdot)$ encodes the desired measure of violation (e.g., identity for expected deviation or $\max(0,\cdot)$ for average violation).

As voltage is a smooth function of loads, \textit{square exponential kernel} has been extensively used for PF learning \cite{liu2022kernel,pareek2021framework,xugp1,xugp2}. The \textit{square exponential kernel} is defined as 
\begin{align}\label{eq:se_kernel}
    k(\mathbf{s}^i,\mathbf{s}^j) = \tau^2 \exp\bigg\{ \frac{-\|\mathbf{s}^i-\mathbf{s}^j\|^2}{2\ell^2} \bigg\},
\end{align}
where $\|\cdot\|$ is the Euclidean norm.

\section{Proofs}\label{app:proofs}
\subsection{Proof of Theorem \ref{thm:1}}\begin{proof}
The expression in the theorem can be written as
\begin{align}
     \big |\mathbb{E}&[h_p(\mathbf{s})] - \widehat{\mathbb{E}}[h_m(\mathbf{s})] \big |\nonumber\\& = \big |\mathbb{E}[h_p(\mathbf{s})] - \mathbb{E}[h_m(\mathbf{s})] + \mathbb{E}[h_m(\mathbf{s})] - \widehat{\mathbb{E}}[h_m(\mathbf{s})] \big | \nonumber \\
     & \leq \underbrace{\big |\mathbb{E}[h_p(\mathbf{s})] -\mathbb{E}[h_m(\mathbf{s})] \big |}_{\equiv \, B} + \underbrace{\big |\mathbb{E}[h_m(\mathbf{s})] - \widehat{\mathbb{E}}[h_m(\mathbf{s})] \big |}_{\equiv \, C}\label{eq:A}
\end{align}
Using Jensen's inequality, we can upper bound $B$ as follows
\small
\begin{align}\label{eq:B}
   B\leq \mathbb{E}\big[ |h_p(\mathbf{s})-h_m(\mathbf{s}) |\big] \leq \int_{\mathcal{S}}\big | h_p(\mathbf{u})-h_m(\mathbf{u}) \big | \mathcal{P}(\mathbf{u})d\mathbf{u}
\end{align}
\normalsize
Under the theorem's assumption, $h(\cdot)$ is a Lipschitz continuous function with Lipschitz constant $L$, and $M$ is a uniform bound satisfying $M > h_p(\mathbf{s})$. Using the assumption on the ML model's performance, i.e., $\mathbb{P}\{|V(\mathbf{s}) -\widehat{V}(\mathbf{s})|> \varepsilon_m \} \leq \delta_{\epsilon}$ for $\varepsilon_m > 0$ and $\delta_{\epsilon} \in (0,1)$, we have for any $\varepsilon>0$, 
\begin{align}\label{eq:gp_relation_1}
 &\mathbb{P}\big \{ \big | h_p(\mathbf{s}) -h_m(\mathbf{s}) \big |> L\varepsilon \big | {\mathbf{s}\in\mathcal{S}}\big \} \leq \delta_{\epsilon} \nonumber\\
\Rightarrow &\int_{\mathcal{S}} \mathbbm{1}_{\{| h_p(\mathbf{u}) -h_m(\mathbf{u}) |> L\varepsilon\}} \,d\mathbf{u} \leq \delta_{\epsilon}
\end{align}
Here, $\mathbbm{1}$ is the indicator function (1 if $\{| h_p(\mathbf{u}) -h_m(\mathbf{u}) |\geq L\varepsilon\}$ for $\mathbf{u} \in \mathcal{S}$). Let $\mathcal{S}^+ \subset \mathcal{S}$ such that $\forall \,\mathbf{s} \in \mathcal{S}^+$, $| h_p(\mathbf{s})-h_m(\mathbf{s}) | < L\varepsilon_m$ and $\mathcal{S}^- = \mathcal{S}\setminus\mathcal{S}^+$. Then \eqref{eq:B} can be expressed as
\begin{align}
       &B \leq\hspace{-6pt}  \int_{\mathcal{S}^+}\hspace{-8pt}| h_p(\mathbf{u})-h_m(\mathbf{u}) | \mathcal{P}(\mathbf{u})d\mathbf{u}  + \hspace{-6pt}\int_{\mathcal{S}^-}\hspace{-8pt} | h_p(\mathbf{u})-h_m(\mathbf{u}) | \mathcal{P}(\mathbf{u})d\mathbf{u}\nonumber\\
&~~\leq L \varepsilon_m \int_{\mathcal{S}^+}\mathcal{P}(\mathbf{u}) d\mathbf{u} + 2 M \int_{\mathcal{S}^-} \mathcal{P}(\mathbf{u})d\mathbf{u}\label{ub_1}\\
&\Rightarrow B \leq L\varepsilon_m + 2M\delta \text{~for any $\varepsilon_m>0$} \label{eq:B1}
\end{align} 
Here \eqref{ub_1} follows from definition of $\mathcal{S}^+$, $L$ and $M$, and \eqref{eq:B1} follows from \eqref{eq:gp_relation_1} and definition of $\mathcal{S}^+,\mathcal{S}^-$. Next, through direct application of Hoeffding's inequality on $C$ in \eqref{eq:A}, we get 
\begin{align}\label{eq:C}
    C \,\equiv \,\, \mathbb{P}\Big\{\big |\mathbb{E}[h_m(\mathbf{s})] - \widehat{\mathbb{E}}[h_m(\mathbf{s})] \big | \leq \varepsilon_h \Big \} \geq 1-\beta\quad 
\end{align}
where, $\varepsilon_h = \sqrt{\frac{4M^2\log{(2/\beta)}}{N}}$ and $N$ is the number of sample evaluations. Using \eqref{eq:B1} and \eqref{eq:C} in \eqref{eq:A} proves the theorem.
\end{proof}

\subsection{Proof of Theorem \ref{thm:2}}
\begin{proof}
\begin{align*}
\mathbb{P}\big \{h_p(\mathbf{s}) > 0  \big \}
&\leq 
\mathbb{P}\big \{h_p(\mathbf{s}) > 0 \, \, \cap \, \, h_p(\mathbf{s}) > h_m(\mathbf{s}) \big \} \\
&\quad + \mathbb{P}\big \{h_p(\mathbf{s}) > 0 \, \cap \, h_p(\mathbf{s}) \leq h_m(\mathbf{s}) \big \} \\
&\leq 
\mathbb{P}\big \{ h_p(\mathbf{s}) > h_m(\mathbf{s})\big \} \\ 
& \quad + \mathbb{P}\big \{ h_p(\mathbf{s}) > 0 \, \cap \, h_p(\mathbf{s}) \leq h_m(\mathbf{s})\big \}
\end{align*}
as joint probability is always less than individual probabilities. 
$\mathbb{P}\big\{\big |V(\mathbf{s}^i) -\widehat{V}(\mathbf{s}^i) \big | > \varepsilon_{m}\big\} \leq \delta $.
Here, by GP confidence bound $\mathbb{P}\big \{ h_p(\mathbf{s}) - h_m(\mathbf{s}) > \varepsilon \big \} \leq \delta(\kappa)$ for any $\varepsilon \geq 0$, where $\kappa$ decides the confidence level like $\kappa = 3$ for 99.7\% success or $\delta(\kappa)= 0.003$. Thus, 
$$\mathbb{P}\big \{ h_p(\mathbf{s}) > 0    \big \}
\leq 
\delta(\kappa) +
\mathbb{P}\big \{ h_p(\mathbf{s}) > 0 \, \cap \, h_p(\mathbf{s}) \leq h_m(\mathbf{s})\big \}
$$
Now breaking the joint probability as conditional probability $\mathbb{P}(A\, \cap \, B) =\mathbb{P}(A|B) \mathbb{P}(B)$  
$$
\leq 
\delta(\kappa) +
\mathbb{P}\big \{ h_p(\mathbf{s}) > 0| h_p(\mathbf{s}) \leq h_m(\mathbf{s})\big \}\mathbb{P}\big \{ h_p(\mathbf{s}) \leq h_m(\mathbf{s})\big \}
$$
If $h_p(\mathbf{s}) \leq h_m(\mathbf{s})$ and  $h_p(\mathbf{s}) > 0$ then $h_m(\mathbf{s}) > 0$. Thus,
$$
\leq 
\delta(\kappa) +
\mathbb{P}\big \{ V^i_g > 0| h_p(\mathbf{s}) \leq h_m(\mathbf{s})\big \}\mathbb{P}\big \{ h_p(\mathbf{s}) \leq h_m(\mathbf{s})\big \}
$$
Converting conditional into joint probability 
$$\mathbb{P}\big \{ h_p(\mathbf{s}) > 0    \big \}
\leq 
\delta(\kappa)+
\mathbb{P}\big \{h_m(\mathbf{s}) > 0\, \cap \, h_p(\mathbf{s}) \leq h_m(\mathbf{s})\big \}
$$
Again, as joint probability is always less than individual probabilities 
$$\mathbb{P}\big \{ h_p(\mathbf{s}) > 0    \big \}
\leq 
\delta(\kappa) +
\mathbb{P}\big \{ h_m(\mathbf{s}) > 0\big \}
$$
Applying Hoffding's inequality to GP-based probability estimation with $ \varepsilon = \sqrt{{\log(1/\beta)}/{2N}}$
$$
\mathbb{P}\Big \{ \mathbb{P}\big \{ h_m(\mathbf{s}) > 0\big \} -\widehat{\mathbb{P}}\big \{ h_m(\mathbf{s})> 0\big \} \leq \varepsilon \Big \} \geq 1-  e^{-2N\varepsilon^2}
$$
$$
\text{Or}\quad \mathbb{P}\Big \{ \mathbb{P}\big \{ h_m(\mathbf{s}) > 0\big \}\leq \varepsilon +\widehat{\mathbb{P}}\big \{ h_m(\mathbf{s}) > 0\big \} \Big \} \geq 1-  \beta
$$
Thus, 
$$
\mathbb{P}\big \{ h_p(\mathbf{s}) \geq 0    \big \}
\leq  \delta(\kappa) + \varepsilon +\widehat{\mathbb{P}}\big \{ h_m(\mathbf{s}) \geq 0\big \} $$ \text{with confidence}\,  $1-\beta$

\end{proof}

\end{document}